\documentclass[letterpaper,showpacs,preprintnumbers,amsmath,amssymb,nofootinbib,notitlepage]{revtex4-1}
\usepackage{graphicx}
\usepackage[usenames,dvipsnames]{color}
\usepackage[colorlinks,citecolor=blue]{hyperref}
\usepackage{amsmath}
\usepackage{bbm}
\usepackage{amsfonts}
\usepackage{amssymb}
\usepackage{latexsym}
\usepackage{graphicx}
\usepackage[english]{babel}
\usepackage{multirow}
\usepackage{float}
\usepackage{url}
\usepackage{slashed}
\usepackage{xcolor}
\usepackage{enumerate}
\usepackage{makecell}
\usepackage{diagbox}
\usepackage{makecell}

\newcommand{\be}{\begin{equation}}
\newcommand{\ee}{\end{equation}}
\newcommand{\ba}{\begin{array}}
\newcommand{\ea}{\end{array}}
\newcommand{\bea}{\begin{eqnarray}}
\newcommand{\eea}{\end{eqnarray}}

\newcommand{\ttb}{t\bar t }
\newcommand{\jpsi}{J/\psi}
\newcommand{\ttjpsi}{t\bar t J/\psi}
\newcommand{\ppttjpsi}{p p \to t \bar t+J/\psi}

\newcommand{\GeV}{\,{\rm GeV}}

\newcommand{\nn}{\nonumber}

\usepackage{ulem,fancyvrb}
\usepackage{xcolor}

\begin{document}
\title{Prompt $\jpsi$ production in associated with top quark pair at the LHC }

\author{Gang Li$^1$~}

\author{Xue-An Pan$^1$~}

\author{Mao Song$^1$~}

\author{Yu Zhang $^{2,1}$~} \email{dayu@nju.edu.cn; dayu@ahu.edu.cn}
\affiliation{
$^1$ School of Physics and Materials Science, Anhui University, Hefei 230039,China \\
$^2$ Institutes of Physical Science and Information Technology,
Anhui University, Hefei 230601, China }

\begin{abstract}

In this work, we investigate the prompt $\jpsi$ production in associated with top
quark pair to leading order in the nonrelativistic QCD 
factorization formalism at the LHC with $\sqrt{s} =13$ TeV.
In addition to the contribution from direct $\jpsi$ production,
we also include the indirect contribution from the 
directly produced heavier charmmonia $\chi_{cJ}$ and $\psi^\prime$.
We present the numerical results for the total and differential 
cross sections and find that  the $\sideset{^3}{^{(8)}_1}{\mathop{{S}}}$ states give
the dominant contributions.
The prompt $\ttb\jpsi$ signatures
at the LHC are analyzed in
the  tetralepton channel $pp\to (t\to W^+(\ell^+\nu)b)
(\bar t \to W^-(\ell^- \bar \nu)\bar b) (\jpsi\to\mu^+\mu^-)$
and trilepton channel $pp\to (t\to W(q  q^\prime)b) ( t \to W(\ell  \nu) b) (\jpsi\to\mu^+\mu^-)$,
with the $\jpsi$ mesons decaying into muon pair, and the top quarks decaying 
leptonically or hadronically.  
We find that $\ttjpsi$ proudction can be potentially detected at the LHC,
whose measurement is useful to test the heavy quarkonium production mechanism.

\end{abstract}

\maketitle

\section{Introduction}\label{sec:introduction}

The nonrelativistic QCD (NRQCD) \cite{Bodwin:1994jh} provides a rigorous theory 
to study heavy-quarkonium physics, where the production 
cross section and decay rates can be divided into 
short-distance and long-distance parts. 
The short-distance coefficients are process-dependent 
which can be 
calculated perturbatively in QCD as expansions 
in the strong-coupling constant $\alpha_s$.
The long-distance matrix elements (LDMEs)
are process-independent and univeral which
are governed by nonperturbative QCD dynamics
and can be extracted from experiments.
The relative importance of  the LDMEs can
be estimated by means of the velocity-scaling rules \cite{Lepage:1992tx};
i.e., the LDMEs can be classified with
a definite power of
the realative velocity $v$ of heavy 
quarks in the bound state within the limit $v\ll1$.
As a result, the theoretical predictions can
be experessed as a double series in $\alpha_s$ and $v$.

The charmonium $\jpsi$ associated-production channels 
are very important to test the heavy
quarkonium physics. 
At the hadron colliders, such as LHC and Tevatron, there already exist data
for $\jpsi+\jpsi$ by LHCb \cite{Aaij:2011yc,Aaij:2016bqq},
D0 \cite{Abazov:2014qba}, CMS \cite{Khachatryan:2014iia} and 
ATLAS \cite{Aaboud:2016fzt}, 
$\jpsi+Z$ by ATLAS \cite{Aad:2014kba}, $\jpsi+W$ by ATLAS \cite{Aad:2014rua}, 
$\jpsi$ + charm \cite{Aaij:2012dz} by LHCb.
With the future projected LHC luminosities, more data can be available.
On the theoretical side, many efforts have been made for
the associated-hadroproduction channels,
$\jpsi+\jpsi$ \cite{Sun:2014gca,Likhoded:2016zmk}, 
$\jpsi+Z$ \cite{Kniehl:2002wd,Mao:2011kf,Gong:2012ah},
$\jpsi+W$ \cite{Kniehl:2002wd,Li:2010hc}, 
$\jpsi+\gamma$ \cite{Li:2008ym,Lansberg:2009db,Li:2014ava},
$\jpsi$ + charm \cite{Baranov:2006dh,Artoisenet:1900zz,Qiao:2002nd},
$\jpsi$ + bottom \cite{Gang:2012js},
where some can be know up to next-to-leading-oder (NLO) accuracy
for the short-distance coefficient.

Top quark as the heaviest particle in the standard model (SM),
play an important role in deciphering
the fundamental interactions,
and its studies are currently driven
the Large Hadron Collider (LHC) experiments.
At the LHC, top quarks produced in pairs and 
associated with bosons through
strong interactions, have started to be accessible,
like for example $\ttb\gamma$ \cite{Aad:2015uwa,Aaboud:2017era,Sirunyan:2017iyh},
$\ttb W/Z$ \cite{Aad:2015eua,Aaboud:2016xve,Khachatryan:2015sha,Sirunyan:2017uzs}, and
$\ttb H$ \cite{Aaboud:2018urx,Sirunyan:2018hoz}.
With the running of the LHC, more $\ttb$ associated produciton channels
may can be discovered, such as $\ttb$ production in associated with 
heavy quarkonium which is helpful to understand the heavy quarkonium 
production mechanism and deepen our understanding of the strong interaction.

In this paper, we plan to study the prompt $\jpsi$ +  $\ttb$ production at the LHC
in the NRQCD factorization
formalism to leading-order (LO).
 Prompt $\jpsi$ candidates that can also be produced indirectly via
radiative or hadronic decays of heavier charmonium states, such as 
$\chi_{cJ}\to\jpsi+\gamma$ and $\psi^\prime\to\jpsi+X$,
are not distinguished from directly-produced $\jpsi$.
The respective decay branching ratios are \cite{Tanabashi:2018oca}
${\rm Br}(\chi_{c0}\to\jpsi+\gamma)=(1.40\pm 0.05) \%$,
${\rm Br}(\chi_{c1}\to\jpsi+\gamma)=(34.3\pm 1.0) \%$,
${\rm Br}(\chi_{c2}\to\jpsi+\gamma)=(19.0\pm 0.5) \%$,
and ${\rm Br}(\psi^\prime\to\jpsi+X)=(61.4\pm 0.6) \%$.
Contributions to the total $\ttjpsi$ production rate can
also come from the process of $\ttb+b$-hadrons 
with weak decays of $b\to \jpsi+X$, which can be
separated easily in the detectors. Therefore, we don't consider 
the $\jpsi$ production through $b$-hadron decay here.

This paper is organized as follows.
In Sec. \ref{sec:the-details-of-the-calculation-}, we describe the details 
of the calculation strategies. 
In Sec. \ref{sec:the-results-at-the-the-production-level}, 
we present the numerical results of the total 
and differential cross sections for $\ttjpsi$ 
at the produciton level.
In Sec \ref{sec:signatures}, we analyse 
$\ttjpsi$ signatures at the LHC with the decay of the top
quarks and $\jpsi$.
Discussion and summary are given in Sec. \ref{sec:discussion-and-summary}.

\section{The Details of the calculation }\label{sec:the-details-of-the-calculation-}
In this section, we present the details of
the calculation for the process $\ppttjpsi$
 of prompt $J/\psi$ production
associated with top pair at the LHC in the NRQCD factorization
formalism to LO. 
The cross section for the  direct production of charmonium state $\cal Q$ assiciated with top quark pair
$pp\to \ttb + {\cal Q}$ can be expressed as
\bea
\sigma(pp\to \ttb + {\cal Q})=\int dx_1 dx_2 \sum_{n}\langle {\cal O}^{{\cal Q}}[n]\rangle
\hat{\sigma}(gg\to \ttb+c\bar c[n])[G_{g/A}(x_1,\mu_f)G_{g/B}(x_2,\mu_f)].
\eea
Here $\langle {\cal O}^{{\cal Q}}[n]\rangle$
is the LDME describing the
hadronization of the $c\bar c[n]$ pair
into the observable quarkonium state ${\cal Q}$,
$n$ denotes the $c\bar c$ Fock states contributing at LO 
in $v$, which is specified for ${\cal Q}=\jpsi,\,\psi^\prime,\, \chi_{cJ}$
in Tab. \ref{tab:fockstates}, with  $J=0,1,2$.
From spin symmetry of the heavy quark, the LDMEs at LO in $v$ have the multiplicity relations
\bea
\left\langle {\mathcal{O}^{\jpsi,\psi^\prime}[\sideset{^3}{^{(8)}_J}{\mathop{{P}}}]}\right\rangle &=&
(2J+1)
\left\langle {\mathcal{O}^{\jpsi,\psi^\prime}[\sideset{^3}{^{(8)}_0}{\mathop{{P}}}]}\right\rangle, \nn \\
\left\langle {\mathcal{O}^{\chi_{cJ}}[\sideset{^3}{^{(8)}_1}{\mathop{{S}}}]}\right\rangle &=&
(2J+1)
\left\langle {\mathcal{O}^{\chi_{c0}}[\sideset{^3}{^{(8)}_1}{\mathop{{S}}}]}\right\rangle, \\
\left\langle {\mathcal{O}^{\chi_{cJ}}[\sideset{^3}{^{(1)}_J}{\mathop{{P}}}]}\right\rangle &=&
(2J+1)
\left\langle {\mathcal{O}^{\chi_{c0}}[\sideset{^3}{^{(1)}_0}{\mathop{{P}}}]}\right\rangle. \nn
\eea
$G_{g/A,B}(x,\mu_f)$ represent the distribution
function at the scale $\mu_f$ of gluon
which carries the momentum fraction $x$ of
the proton $A$ or $B$.

\begin{table}[h]
\begin{center}
\begin{tabular}{c|c|c}
\hline
\hline
k  & $\jpsi$, $\psi^\prime$  & $\chi_{c,J}$  \\
\hline
3  & $^3S_1^{(1)}$ & -- \\
5  & --  & $^3P_J^{(1)}$,$^3S_1^{(8)}$ \\
7  &  $^1S_0^{(8)}$, $^3S_1^{(8)}$, $^3P_J^{(8)}$ & -- \\
\hline
\hline
\end{tabular}
\caption{Values of $k$ in the velocity-scaling rule
$\langle {\cal O^{Q}}[n]\rangle \propto v^k$
for the leading $c\bar c$ Fock states $n$ pertinent to $\cal Q$.
}
\label{tab:fockstates}
\end{center}
\end{table}

\begin{figure}[htbp]
\vspace{0.2cm}
\centering
\includegraphics[scale=0.9]{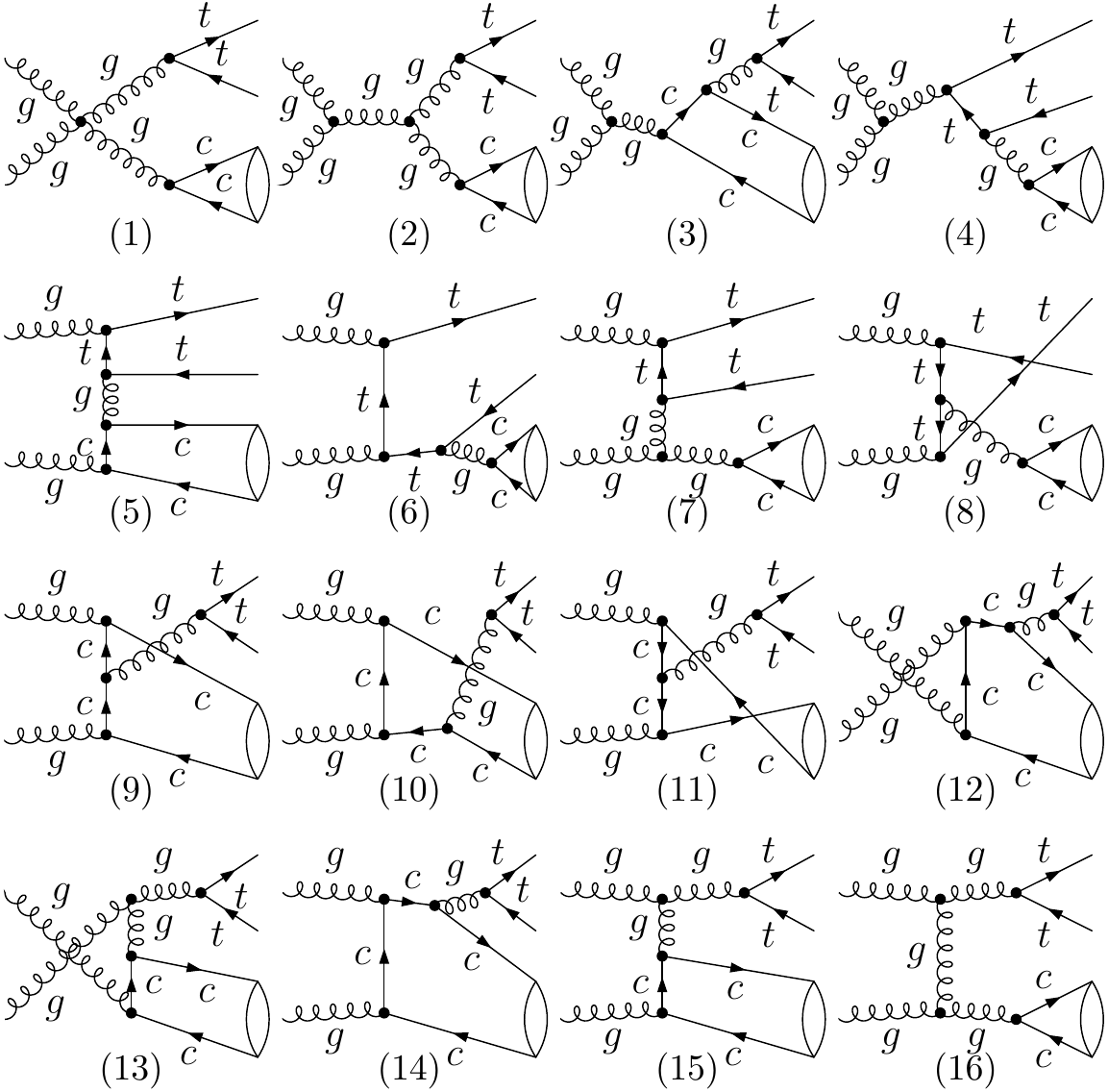}
\caption
{Respresentative diagrams at LO for the partonic process $gg\to \ttb+c\bar c[n]$. }
\label{fig:lofd}
\end{figure}

The $\hat{\sigma}(gg\to \ttb+c\bar c[n])$  decribes
the short-distance cross section for the partonic
process $g (p_1) +g (p_2)\to t(p_3) + \bar t(p_4)+c\bar c[n] (p_5)$ of a $c\bar c$ pair
in a Fock state $n$, and is calculated from the
amplitudes using certain projectors onto the 
usual QCD amplitudes for open $c\bar c$ production.
The amplitude for $gg\to \ttb+c\bar c[n]$ involves 36 Feynman diagrams, which are drawn in 
Fig. \ref{fig:lofd} representatively.

The LO short-distance cross section can be obtained by performing the
integration over the phase space expressed as below,
\be
\hat{\sigma}(gg\to \ttb+c\bar c[n])=\frac{(2\pi)^4}{2\hat{s}
 N_{col}N_{pol}} \int
\overline{\sum}|{\cal A}_{Q\bar{Q} [n]}|^2 d\Omega_3.
\ee
The summation is taken over the spins and colors of initial and
final states, and the bar over the summation denotes averaging over
the spins and colors of initial partons.
The Mandelstam variable $\hat{s}=(p_1+p_2)^2$,
is the partonic center-of-mass system energy.
$N_{col}$ and $N_{pol}$ refer to the numbers of color and polarization
state of $c\bar c[n]$.
The three-body phase space element can be defined as
\be
d\Omega_3=\delta^{(4)}\left(p_1+p_2-\sum_{i=3}^{5} p_i\right)
\prod_{j=3}^{5} \frac{d^3 \vec{p}_j}{(2\pi)^3 2E_j}.
\ee

${\cal A}$ refers to the QCD amplitude with amputated heavy-quark spinors.
In the notations of Ref. \cite{Petrelli:1997ge}:
\bea
{\cal A}_{Q\bar{Q} [{}^1S_0^{(1/8)} ]} &=& {\rm Tr} \Big[ {\cal
C}_{1/8} \Pi_0 {\cal A} \Big]_{q=0}, \nonumber
\\
{\cal A}_{Q\bar{Q} [ ^3S_1^{(1/8)} ]} &=& {\varepsilon}_{\alpha} {\rm Tr}
\Big[ {\cal C}_{1/8} \Pi_{1}^{\alpha} {\cal A} \Big]_{q=0},
\nonumber
\\
{\cal A}_{Q\bar{Q} [ ^1P_1^{(1/8)} ]} &=& {\varepsilon}_{\alpha }
\frac{d}{dq_{\alpha}} {\rm Tr} \Big[ {\cal C}_{1/8} \Pi_0 {\cal A}
\Big]_{q=0}, \nonumber
\\
{\cal A}_{Q\bar{Q} [ ^3P_J^{(1/8)} ]} &=& {\varepsilon}_{\alpha
\beta}^{(J)} \frac{d}{dq_{\beta}} {\rm Tr} \Big[ {\cal C}_{(1/8)}
\Pi_1^{\alpha} {\cal A} \Big]_{q=0},
\eea
where 
the lower index $q$ denotes the momentum of the
charm-quark in the $c\bar{c}$ rest frame. $\Pi_{0/1}$ are spin
projectors onto the spin singlet and spin triplet states. ${\cal
C}_{1/8}$ are color projectors onto the color-singlet and color-octet
states. ${\varepsilon}_{\alpha}$ and ${\varepsilon}_{\alpha \beta}$
represent the polarization vector and tensor of the $c\bar{c}$
states, respectively.

\section{The results at the the production level}\label{sec:the-results-at-the-the-production-level}

In this section, we present the numerical results for the production of prompt $\jpsi$ mesons
in associated with top pair at the LHC with $\sqrt s = 13$ TeV.
The masses of the charm quark and top quark  are taken as $m_c=1.5$ GeV
and $m_t=172$ GeV.
We use the CTEQ6L1 parton distribution functions \cite{Pumplin:2002vw}
with a one-loop running $\alpha_s$ in the LO calculations.
Our default choice of factorization scale is $\mu_f= m_T^{\cal Q}$,
where $m_T^{\cal Q}=\sqrt{(p_T^{\cal Q})^2+4 m_{c}^2}$ is
the charmina ${\cal Q}$ transverse mass ,
and $p_T^{\cal Q}$ is the charmina ${\cal Q}$ transverse momentum.
We apply the cut
\be
p_T^{\cal Q}> 3\,\GeV
\ee
to the final charmonia, which is defined as the ``basic cut" hereafter.
We use {\sc FeynArts} \cite{Hahn:2000kx} to generate Feynman diagrams and amplitudes
for the partonic process $gg\to \ttb+c\bar c[n]$.
Then the corresponding amplitudes are reduced with {\sc FeynCalc} \cite{Shtabovenko:2016sxi} and 
{\sc FeynCalcFormLink} \cite{Feng:2012tk}. At last the numerical calculations are obtained with 
{\sc FormCalc} \cite{Hahn:1998yk}.

We adopt the LDMEs for direct $\jpsi$ production from \cite{Fleming:1997fq,Kniehl:2001tk} as
\bea
\left\langle {\mathcal{O}^{J/\psi}[\sideset{^3}{^{(1)}_1}{\mathop{{S}}}]}\right\rangle &=& 1.1 \;\mathrm{GeV^3}, \notag \\
\left\langle {\mathcal{O}^{J/\psi}[\sideset{^1}{^{(8)}_0}{\mathop{{S}}}]}\right\rangle &=& 1\times 10^{-2} \;\mathrm{GeV^3}, \notag \\
\left\langle {\mathcal{O}^{J/\psi}[\sideset{^3}{^{(8)}_1}{\mathop{{S}}}]}\right\rangle &=& 1.12\times 10^{-2} \;\mathrm{GeV^3}, \notag \\
\left\langle {\mathcal{O}^{J/\psi}[\sideset{^3}{^{(8)}_0}{\mathop{{P}}}]}\right\rangle &=& 11.25\times 10^{-3} \;\mathrm{GeV^5}.
\eea
The LDMEs for direct $\psi^\prime$ and $\chi_{cJ}$ production and 
are choosen from \cite{Kniehl:2006qq} and \cite{Eichten:1995ch,Ma:2010vd} as
\bea
\left\langle {\mathcal{O}^{\psi^\prime}[\sideset{^3}{^{(1)}_1}{\mathop{{S}}}]}\right\rangle &=& 0.67 \;\mathrm{GeV^3}, \notag \\
\left\langle {\mathcal{O}^{\psi^\prime}[\sideset{^1}{^{(8)}_0}{\mathop{{S}}}]}\right\rangle &=& 5\times 10^{-3} \;\mathrm{GeV^3}, \notag \\
\left\langle {\mathcal{O}^{\psi^\prime}[\sideset{^3}{^{(8)}_1}{\mathop{{S}}}]}\right\rangle &=& 2\times 10^{-3} \;\mathrm{GeV^3}, \notag \\
\left\langle {\mathcal{O}^{\psi^\prime}[\sideset{^3}{^{(8)}_0}{\mathop{{P}}}]}\right\rangle &=& 3.214\times 10^{-3} \;\mathrm{GeV^5},
\eea
and
\bea
\left\langle {\mathcal{O}^{\chi_{c0}}[\sideset{^3}{^{(8)}_1}{\mathop{{S}}}]}\right\rangle &=& 2.2 \times 10^{-3} \;\mathrm{GeV^3}, \notag \\
\left\langle {\mathcal{O}^{\chi_{c0}}[\sideset{^3}{^{(1)}_0}{\mathop{{P}}}]}\right\rangle &=&
\frac{3N_c}{2\pi} \times 0.075 \;\mathrm{GeV^5},
\eea
with $N_c=3$.
In our calculations, the relations of LDMEs with conventions of 
Bodwin-Braaten-Lepage have been considered \cite{Petrelli:1997ge}.

In Tab. \ref{tab:directcs}, we present the cross section of
the direct production of the charmina $\jpsi$, $\psi^\prime$, $\chi_{cJ}$
associated with top quark pair at the 13 TeV LHC.
We list the contribution from each $c\bar c$ Fock state,
respectively. It can be seen that 
the contribution from the Fock state $\sideset{^3}{^{(8)}_1}{\mathop{{S}}}$ 
is dominant for all the direct production of charmina $\jpsi$, $\psi^\prime$, $\chi_{cJ}$
in associated with $\ttb$ at the 13 TeV LHC.  
Especially for $\jpsi$ ($\chi_{c2}$), $\sideset{^3}{^{(8)}_1}{\mathop{{S}}}$  state can contribute about 714 (701) fb 
which is nearly 96.3\% (96.2\%)of the corresponding total cross section.
The summation of cross section for the processes $pp\to \ttb \chi_{cJ}$
with $J=0,1,2$, can reach 1350 fb, which will 
provide abundant and fascinating studies of phenomenology
at the LHC \cite{nextwork}.

We summarize the direct and indirect contributions from the
radiative or hadronic decays of heavier charmonium states
for the prompt $\jpsi$ production
in Tab. \ref{tab:indirectcs}. 
The cross sections of the four residual indirect 
production channels can be obtained approximately
by multiplying the cross sections of
the respective intermediate directly produced 
charmonium states with their decay braching ratios
to $\jpsi$ mesons: 
\bea
\sigma^{\rm indirect}({\rm From\,\, \psi^\prime})=\sigma(pp\to \ttb +{\psi^\prime})\times 
{\rm Br}({\psi^\prime}\to \jpsi + X ) \\
\sigma^{\rm indirect}({\rm From}\,\, \chi_{cJ})=\sigma(pp\to \ttb +\chi_{cJ})\times 
{\rm Br}(\chi_{cJ}\to \jpsi + \gamma ) 
\eea
We can see that the total prompt $\ttjpsi$ produciton rates
at the 13 TeV LHC can reach more than 1100 fb,
of which the indirect contribution accounts for about 34\%.
The indirect contribution mainly comes from the $pp\to\ttb\chi_{c2}$ and 
$pp\to\ttb\chi_{c1}$ channels which can account
for nearly 77\% of the total indirect contribution.
The indirect contribution from $\chi_{c0}$ is less than 1\%, which can be neglected. 

\begin{table}[h]
\begin{center}
\begin{tabular}{c|c|c|c|c|c|c}
\hline
\hline
Fock state & $\sideset{^3}{^{(1)}_1}{\mathop{{S}}}$ & $\sideset{^1}{^{(8)}_0}{\mathop{{S}}}$
& $\sideset{^3}{^{(8)}_1}{\mathop{{S}}}$ & $\sideset{^3}{^{(8)}_J}{\mathop{{P}}}$ &$\sideset{^3}{^{(1)}_J}{\mathop{{P}}}$ & Total \\
\hline
$\sigma(\ppttjpsi)$ & 0.125 & 9.57 & 714.08 & 17.25 & -- & 741.03 \\
\hline 
$\sigma(pp\to \ttb +\psi^\prime)$ & 0.077 & 4.79 & 127.51 & 4.93 &-- & 137.31 \\
\hline
$\sigma(pp\to \ttb +\chi_{c2})$ & --& -- & 701.33 & -- & 27.67 & 729.00 \\
\hline
$\sigma(pp\to \ttb+ \chi_{c1})$  & --& -- & 420.80 & -- & 43.95 & 464.75 \\
\hline
$\sigma(pp\to \ttb+ \chi_{c0})$  & --& -- & 140.27 & -- & 16.23 & 156.50 \\
\hline
\hline
\end{tabular}
\caption{Total cross section (in unit of fb)  of the  direct charmonium production process 
$pp\to\ttb+{\cal Q}$ at the 13 TeV LHC with basic cut.
}
\label{tab:directcs}
\end{center}
\end{table}
\begin{table}[h]
\begin{center}
\begin{tabular}{c|c|c|c|c|c|c|c}
\hline
\hline
Source & From $\psi^\prime$ & From $\chi_{c2}$ & From $\chi_{c1}$ & From $\chi_{c0}$  &  Indirect & Direct  & Prompt\\
\hline
$\sigma$ (fb) & 84.31 & 138.51 & 159.41 & 2.19 & 384.42 & 741.03 & 1125.45  \\
\hline
\hline
\end{tabular}
\caption{Total cross section for $\jpsi$ production  from direct  and indirect contributions at 13 TeV LHC with basic cut.
}
\label{tab:indirectcs}
\end{center}
\end{table}

\begin{figure}[htbp]
\vspace{0.2cm}
\centering
\includegraphics[scale=0.4]{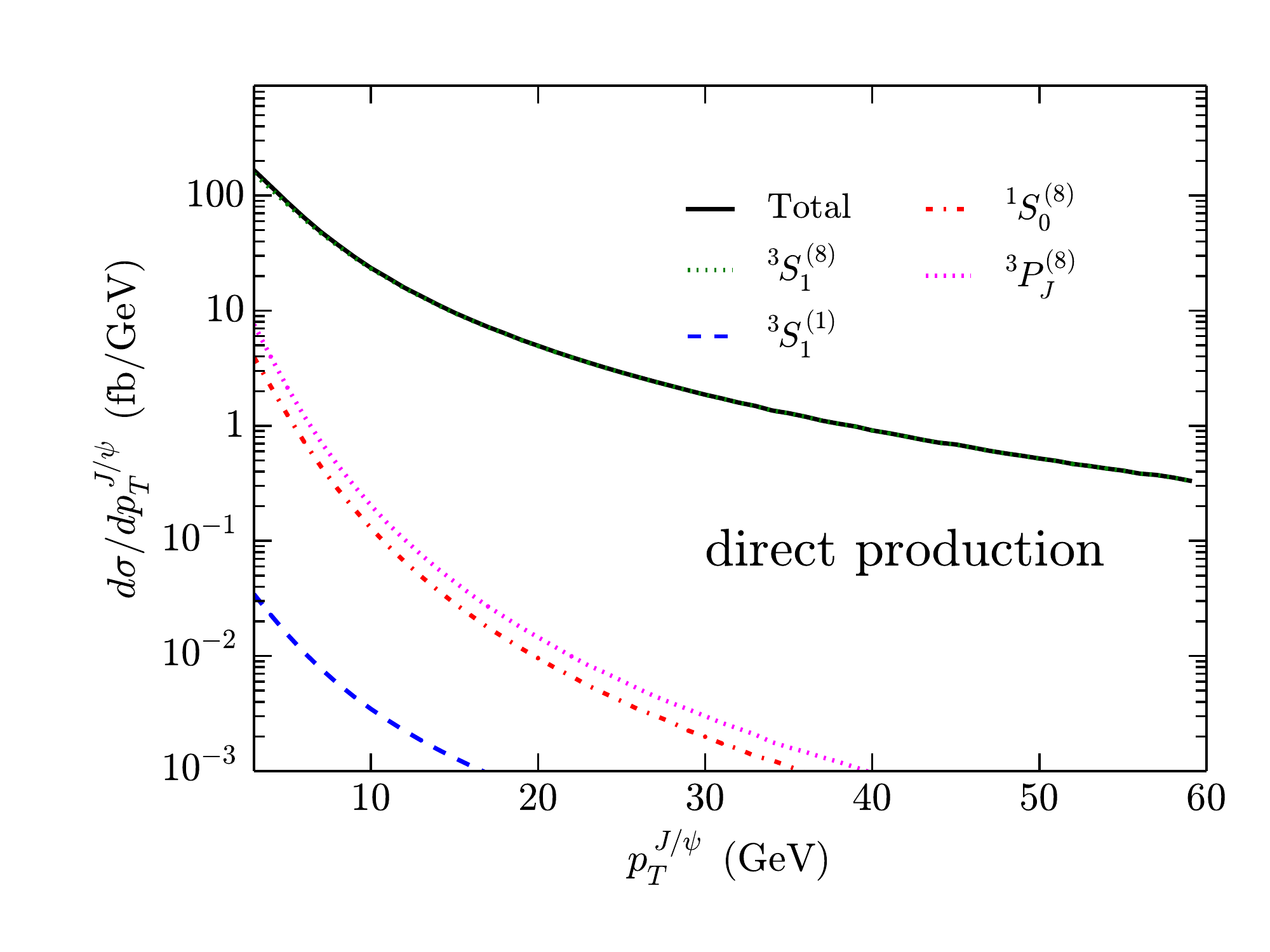}
\includegraphics[scale=0.4]{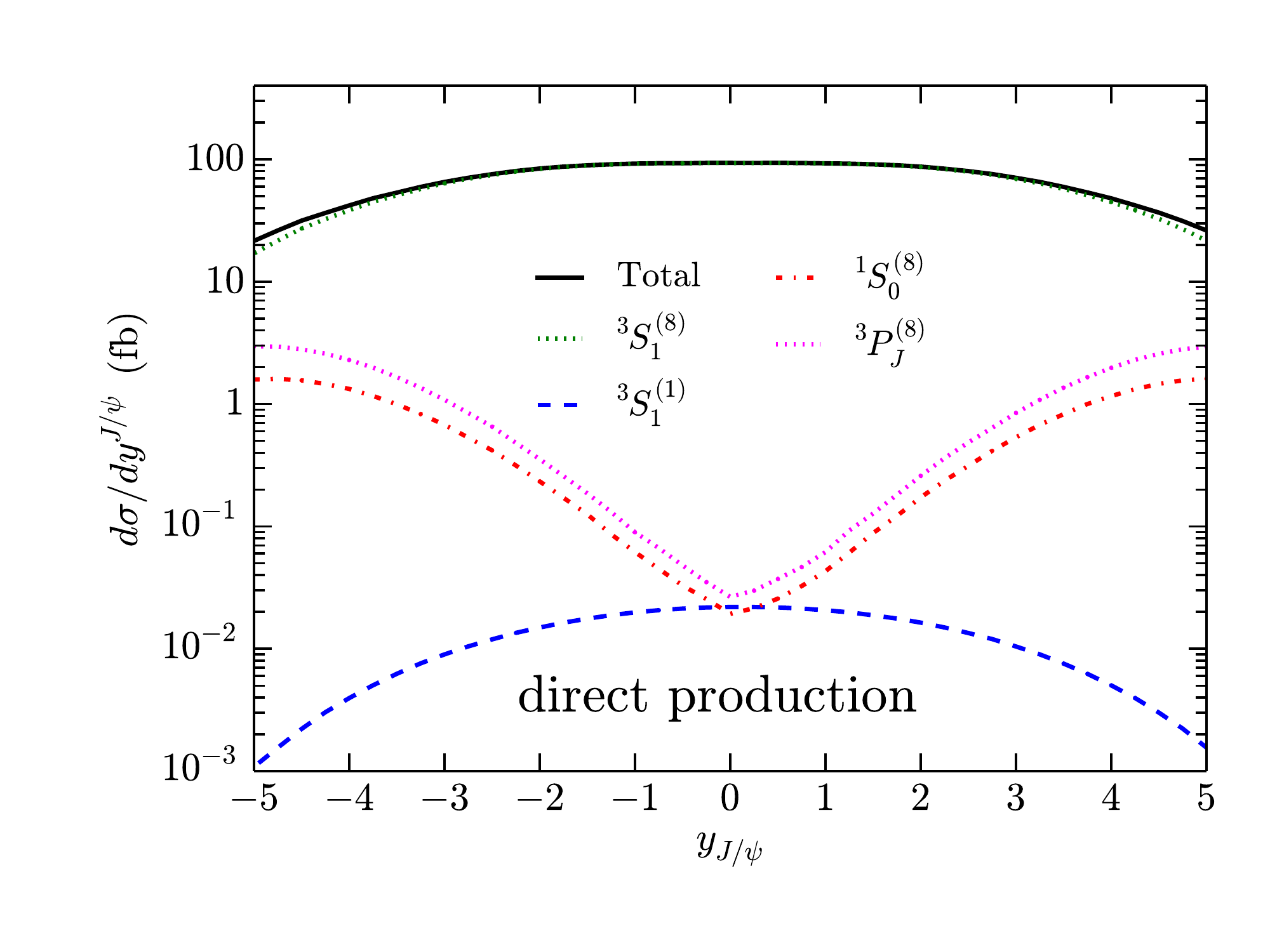}
\caption
{The LO distributions of  
$p_T^{\jpsi}$ and  $y_{\jpsi}$ for the direct $\jpsi$ production process 
$\ppttjpsi$ 
at the 13 TeV LHC. The contributions from Fock states
$\sideset{^3}{^{(1)}_1}{\mathop{{S}}}$, $\sideset{^1}{^{(8)}_0}{\mathop{{S}}}$,
$\sideset{^3}{^{(8)}_1}{\mathop{{S}}}$ and $\sideset{^3}{^{(8)}_J}{\mathop{{P}}}$
are listed respectively. }
\label{fig:direct}
\end{figure}

In Fig. \ref{fig:direct}, we present the distributions of the final $\jpsi$ transverse momentum 
$p_T^{\jpsi}$ and the rapidity $y_{\jpsi}$ for the direct production process 
$\ppttjpsi$ at the 13 TeV LHC.
For camparison, we also show the Fock states
$\sideset{^3}{^{(1)}_1}{\mathop{{S}}}$, $\sideset{^1}{^{(8)}_0}{\mathop{{S}}}$,
$\sideset{^3}{^{(8)}_1}{\mathop{{S}}}$ and $\sideset{^3}{^{(8)}_J}{\mathop{{P}}}$
there. We can see that the $\jpsi$ spectrum of 
all the Fock states steadily decrease with the increment of $p_T^{\jpsi}$.
In the range of $3 \GeV \leq p_T^{\jpsi}\leq 40 \GeV $, 
the $d\sigma/dp_T^{\jpsi}$ is in the range of [0.91, 154.18] fb/GeV.
We can see that the $\sideset{^3}{^{(8)}_1}{\mathop{{S}}}$ state gives
the dominant contribution in the whole plotted region,
and the curves of total and $\sideset{^3}{^{(8)}_1}{\mathop{{S}}}$ state contribution 
for the distribution are almost overlap.
For the $y_{\jpsi}$ distributions,
at the center region $y_{\jpsi}=0$,
the contributions from $\sideset{^3}{^{(1)}_1}{\mathop{{S}}}$ and $\sideset{^3}{^{(8)}_1}{\mathop{{S}}}$
states reach their maximum, while the contributions from $\sideset{^1}{^{(8)}_0}{\mathop{{S}}}$
and $\sideset{^3}{^{(8)}_J}{\mathop{{P}}}$ states reach their minimum.

\begin{figure}[htbp]
\vspace{0.2cm}
\centering
\includegraphics[scale=0.4]{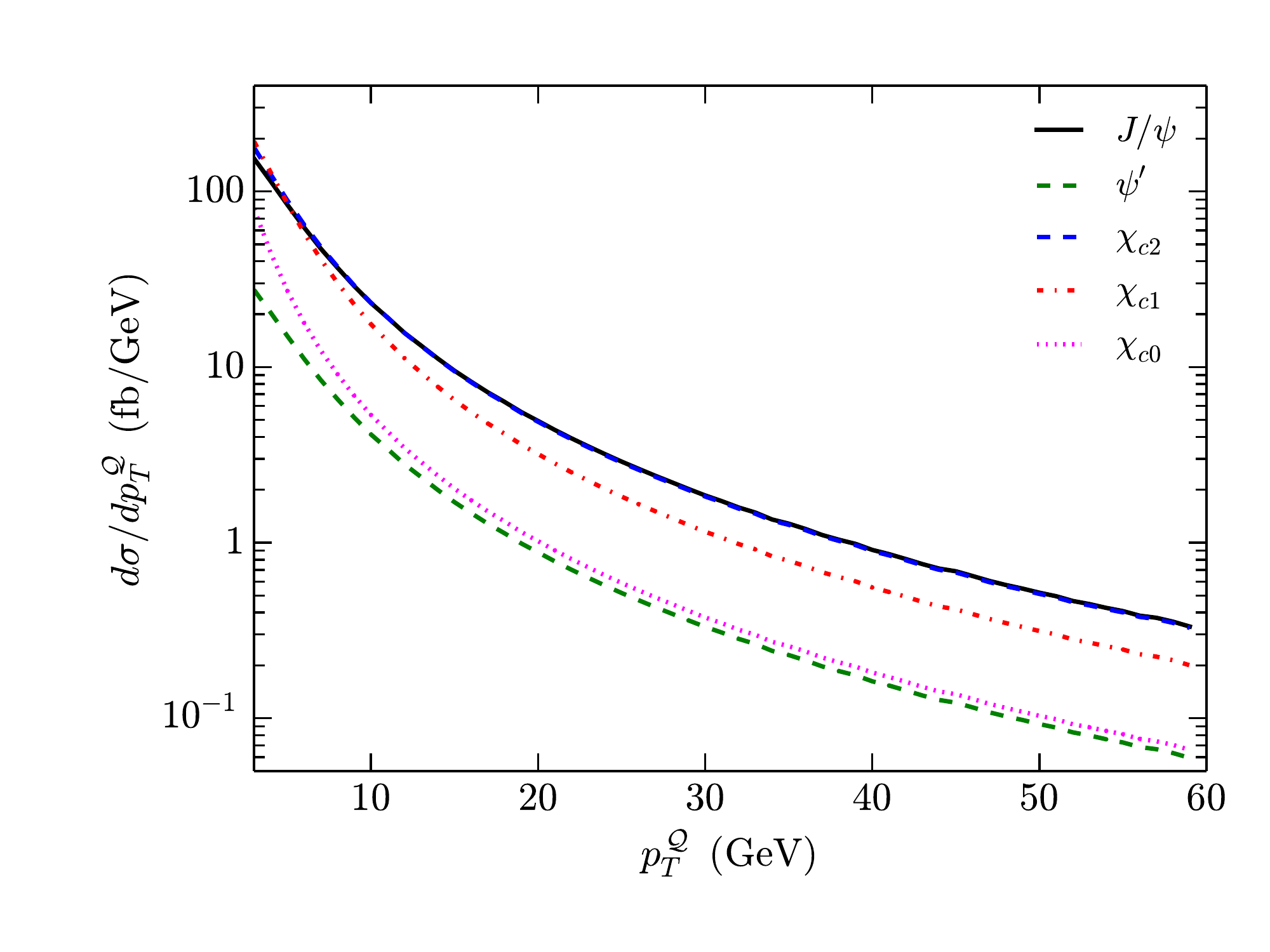}
\includegraphics[scale=0.4]{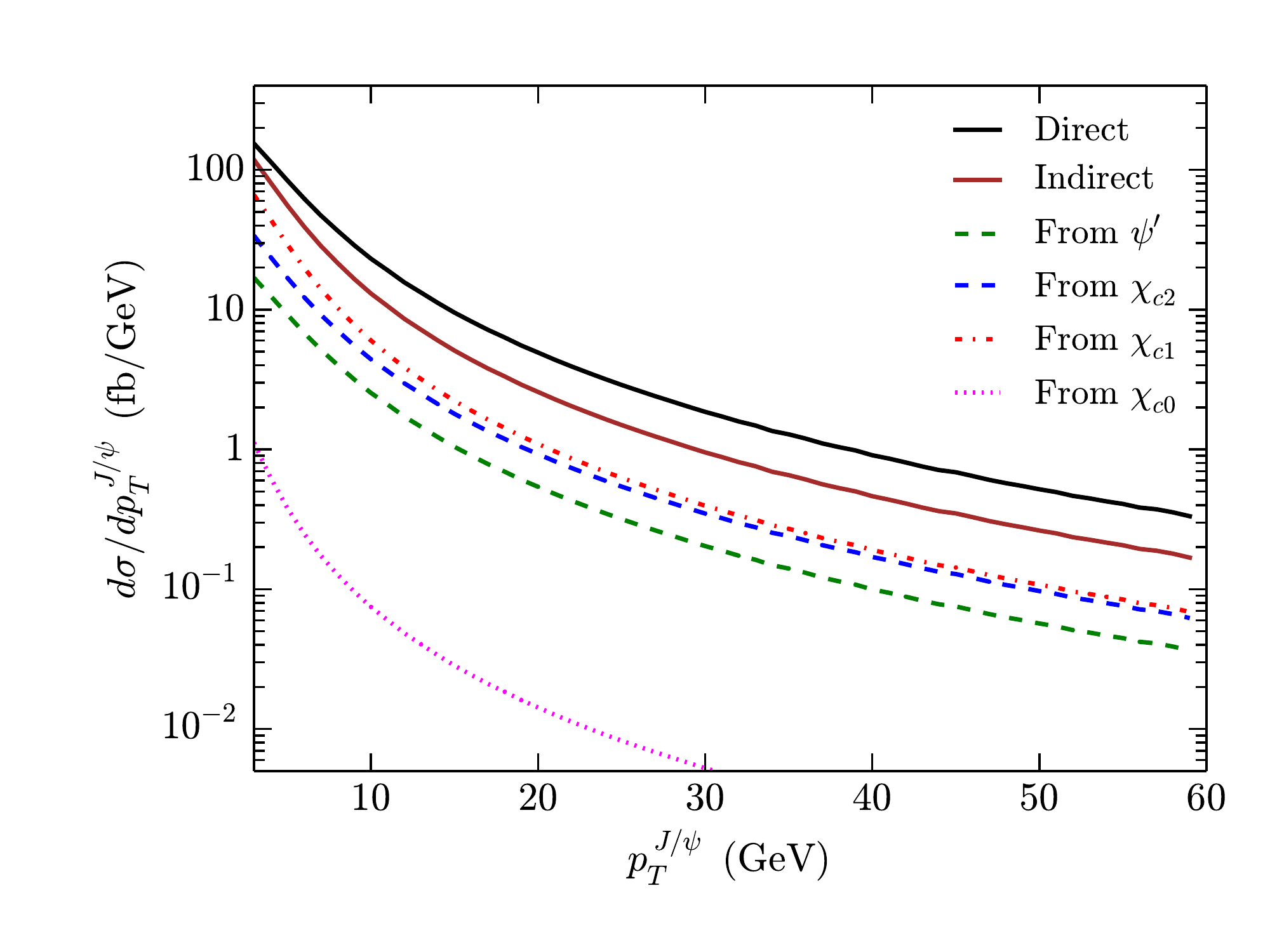}
\caption
{Left: The LO $p_T$ distributions for the direct charmina $\cal Q$ production process 
$pp\to \ttb +{\cal Q}$ 
at the 13 TeV LHC. Right: The contribution from the direct and indirect $\jpsi$ production for 
LO  $p_T^{\jpsi}$ distributions.}
\label{fig:indirect}
\end{figure}

The transverse momentum distributions for the direct production charmina ${\cal Q}$
in the channels $pp\to \ttb +{\cal Q}$ are shown in the Fig. \ref{fig:indirect} on the left.
We also plot the $p_T^{\jpsi}$ spectra from the indirect contributions in 
the Fig. \ref{fig:indirect} on the right.
We can find that, at 13 TeV LHC,  the $p_T$ distribution 
for the direct production  $\chi_{c2}$ and $\chi_{c1}$ are larger 
in the lower $p_T$ region than $\jpsi$.
The behavior of the spectra for the direct and indirect contributions are similar;
and the indirect contribution from $\chi_{c0}$ is much less than other channels in 
the whole plotted $p_T^{\jpsi}$ region.

\section{Analyses of $\ttjpsi$ signatures at the LHC}\label{sec:signatures}

In this section, we analyse the prompt $\jpsi$ production associated
with top quark pair  signatures at the LHC.
We focus on the $\jpsi$ decaying into a pair of  opposite-sign
(OS) muons in $\ttjpsi$ production
\footnote{The $\jpsi\to e^+e^-$ channel is not used because the ATLAS detector
has poor signal effciency for the $\jpsi$ relevant momentum range \cite{Bertsche:2016skf}.
}, which presents as irreducible background the $t\bar t \mu^+\mu^-$ production.
The events for the background $pp\to \ttb \mu^+\mu^-$ are simulated with {\sc Madgraph} 
\cite{Alwall:2014hca}.
The $\jpsi\to \mu^+\mu^-$ decay is implemented with
the narrow width approximation (NWA) method,
and event weights are rescaled by the branching ratio
BR$(\jpsi\to\mu^+\mu^-)=5.961\%$, which is taken from \cite{Tanabashi:2018oca}.

\subsection{Stable top quarks}
\label{sec:stable}
We begin our discussion without including top quark decays.
For the final muons of the signal and background processes,
the following cuts are applied:
\bea
 p_{T}^{\mu^\pm} > 2.5\GeV, \,\,\,|\eta_{\mu^\pm}|<2.3,\, \nn \\
 |m(\mu^+\mu^-)-m_{\jpsi}|<0.5 \GeV,
 \label{eq:nodecay-cut1}
\eea
where $p_{T}^{\mu^\pm}$ and $\eta_{\mu^\pm}$ are the transverse momentum
and pseudorapidity of the final muon and $m(\mu^+\mu^-)$
denotes the invariant mass of the final muon pair.
Using the cuts for $\jpsi$ probed in the ATLAS experiment,
additional cuts are also applied to the final reconstructed
$\jpsi$ mesons \cite{Aad:2014rua,Aad:2014kba}:
\bea
 p_{T}^{\jpsi} > 8.5\GeV, \,\,\,|y_{\jpsi}|<2.1,
 \label{eq:nodecay-cut2}
\eea
where
$p_{T}^{\jpsi}$ and $y_{\jpsi}$ are the transverse momentum rapidity
of the $\jpsi$ mesons reconstructed from the muon pair.

\begin{figure}[htbp]
	\vspace{0.2cm}
	\centering
	\includegraphics[scale=0.34]{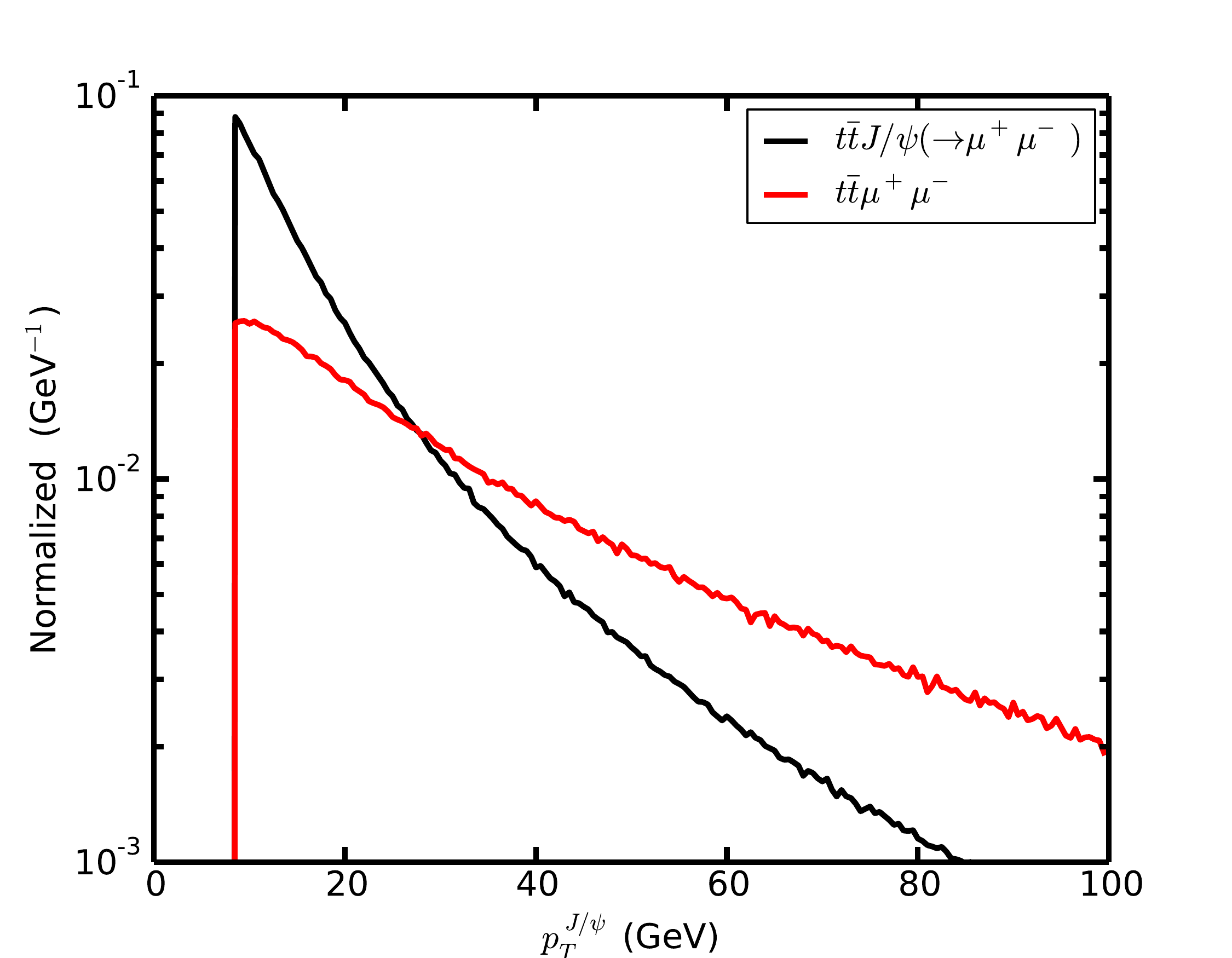}
	\includegraphics[scale=0.34]{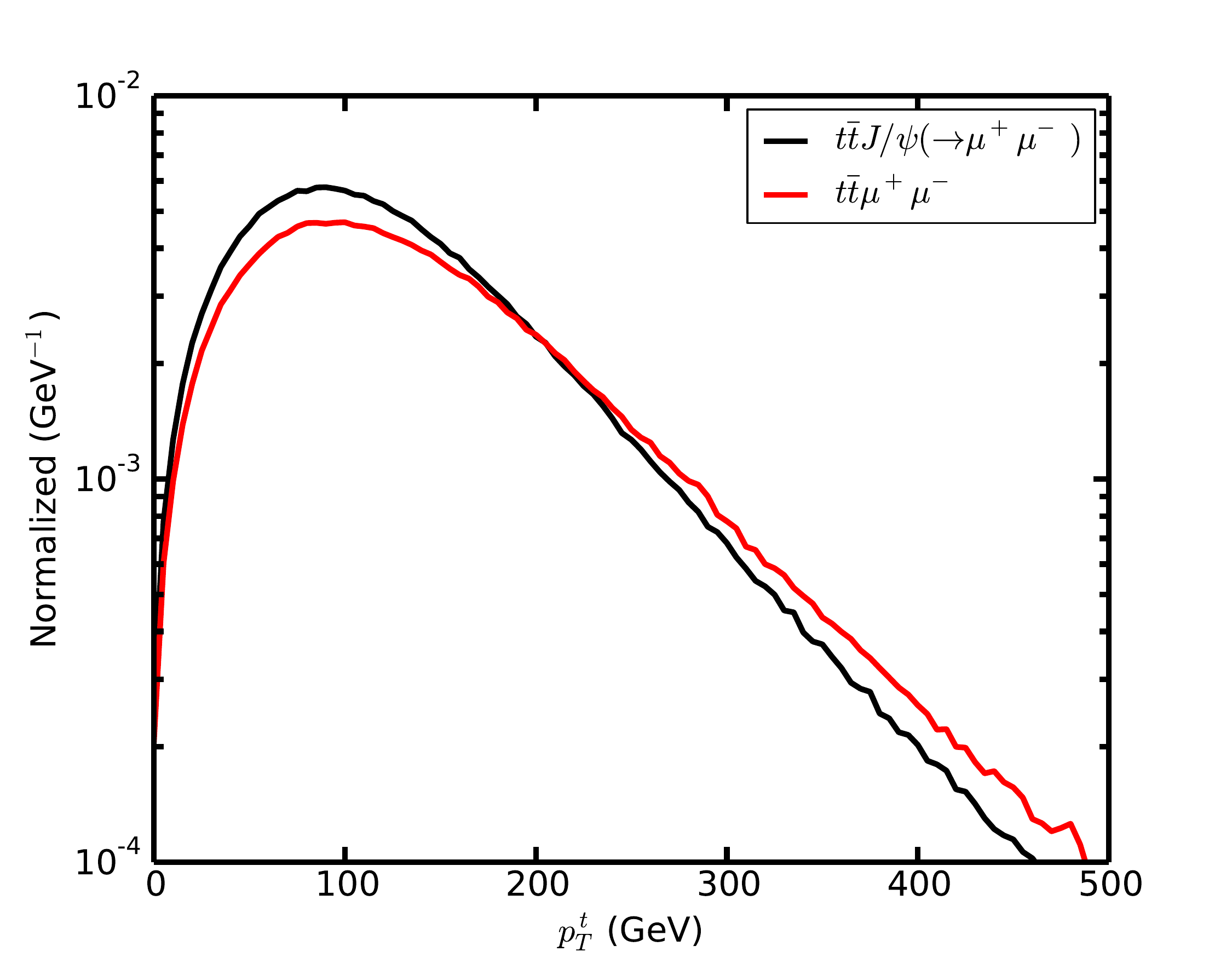}
	\includegraphics[scale=0.34]{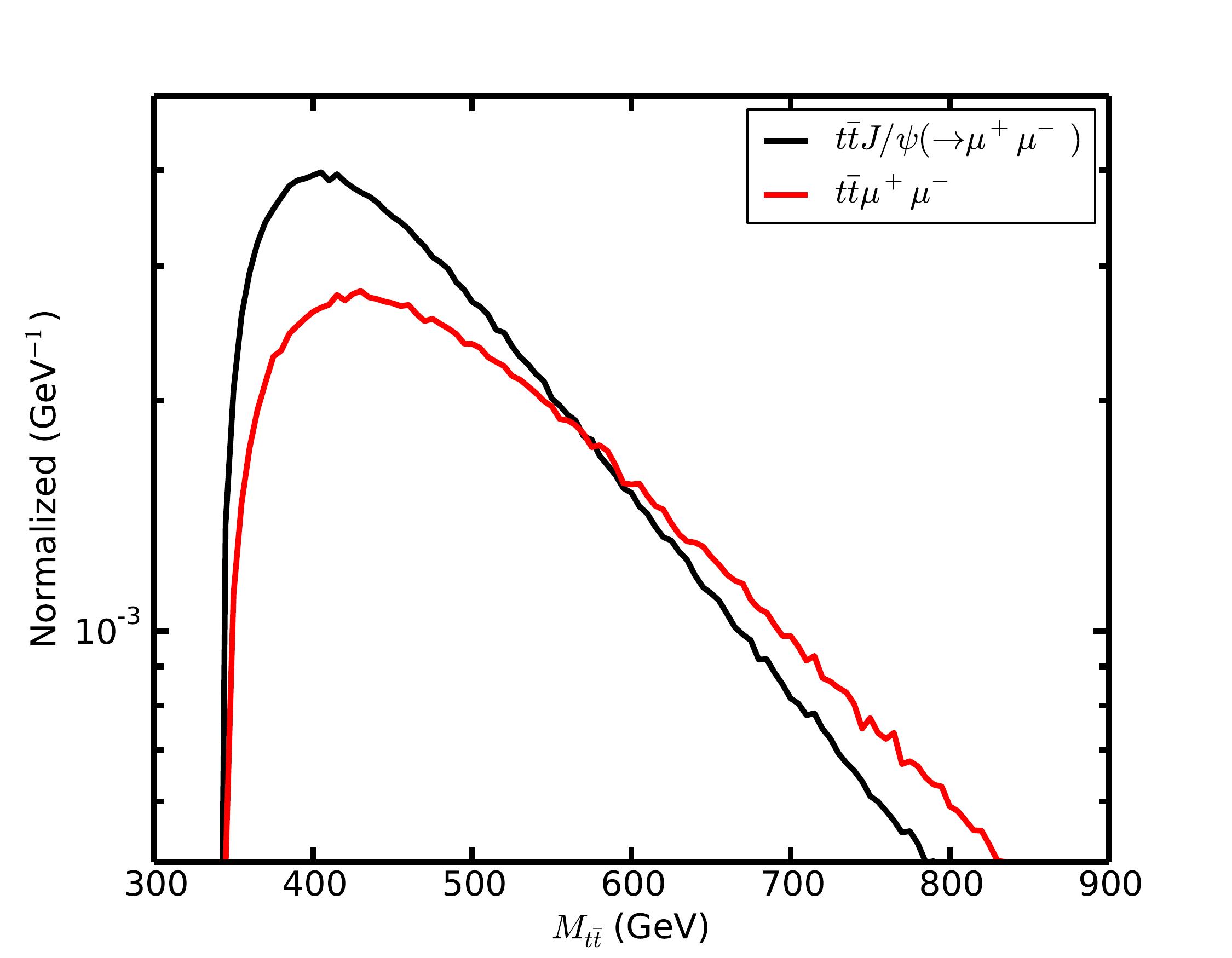}
	\caption
	{The normalized distributions
	of $p_T^t$ and $p_T^{\jpsi}$
	and the top quark pair
	invariant masses $M_{t\bar t}$ for the prompt 
	$\ttb\jpsi(\to \mu^+\mu^-)$ and the background $\ttb\mu^+\mu^-$
	at the 13 TeV LHC.}
	\label{fig:nodecay}
\end{figure}

In Fig. \ref{fig:nodecay},
we display the normalized distributions
of the transverse momentum of the final top ($p_T^t$) and the final
reconstructed $\jpsi$ ($p_T^{\jpsi}$)
and the top quark pair
invariant masses $M_{t\bar t}$ for the prompt 
$\ttb\jpsi(\to \mu^+\mu^-)$ and the background $\ttb\mu^+\mu^-$. We can see that
the signal decrease faster than the background with
increasing $p_T^{\jpsi}$, $M_{t\bar t}$ and $p_T^t$.
In order to increase the significance
${\cal S} = S/\sqrt{B}$, where $S$ ($B$) is the
number of signal (background) events, 
we investigate the $p_T^{\jpsi}$, $M_{t\bar t}$ and $p_T^t$ cuts effects,
and find  
that when
\be
p_{T}^{\jpsi} < 30 \GeV,
\label{eq:nodecay-cut3}
\ee
the significance can be maximized based on the cuts (\ref{eq:nodecay-cut1}) and (\ref{eq:nodecay-cut2}).

In Tab. \ref{tab:nodecay}, we present the total cross sections
for $\ttb$ associated with the direct and indirect production $\jpsi$ 
mesons including
$\jpsi$ mesons decay to muon pair, and the SM background $pp\to\ttb \mu^+\mu^-$,
after each cut above at  13 TeV LHC,
In the last column, the corresponding prompt significance 
${\cal S}={\sigma^{\rm signal}}\sqrt{\rm L}/\sqrt{\sigma^{\rm bkg}} $ is also
given with the luminosity L=100 fb$^{-1}$.
After all the selection cuts, we can see that with L=100 fb$^{-1}$ the total 
number of events for the prompt $\ttjpsi$ production
with the subsequent decay $\jpsi\to\mu^+\mu^-$ can account for more than 600
and the corresponding significance can reach more than 100.
The reason for the decline of the significance
after the cut (\ref{eq:nodecay-cut2}) is that the contributions from the lower $p_T^{\jpsi}$
region dominate, which can be seen in the Fig. \ref{fig:direct}. Therefore, we suggest that the $p_T^{\jpsi}$
should be measured as lower as the experiment can to get larger significance and more
signal events.

\begin{table}[h]
\begin{center}
\begin{tabular}{|c|c|ccccc|c|c|}
\hline\hline
\multirow{2}*{\diagbox{Cut}{Contribution}}  & \multirow{2}*{Direct} & & & Indirect & & &
\multirow{2}*{Background} & \multirow{2}*{$\cal S$}
\\
\cline{3-7}
& 
& From $ \psi(2S)$
& From $\chi_{c2}$
& From $\chi_{c1}$
& From $\chi_{c0}$ 
& Total Indirect 
&  & \\
\hline

Cut (\ref{eq:nodecay-cut1}) & 12.56  & 1.38	& 2.37 	& 2.65	
& 0.036	& 6.44 	  & 0.82	& 209.8 \\

Cut (\ref{eq:nodecay-cut2}) & 5.62  & 0.62	& 1.05	& 1.145	
& 0.015 &	2.83  	& 0.74 & 98.2 \\  %
Cut (\ref{eq:nodecay-cut3}) & 4.25  & 0.47	& 0.79 	& 0.86	
& 0.01	& 2.13	& 0.34	&109.4 \\  %
\hline\hline
\end{tabular}
\caption{The cross sections (in unit of fb) for the signal from the direct and prompt $\jpsi$
and background after each cut at the 13 TeV LHC. The significance is given with L=100 fb$^{-1}$.}
\label{tab:nodecay}
\end{center}
\end{table}

\subsection{Unstable top quarks}
Then we continue our discussion including top quark decays.
For both the signal and background,
the top quarks also decay using NWA method with BR($t\to Wb$) = 100\%;
both $W$ bosons, resulting from the top-quark decays,
are required to decay leptonically or hadronically with 
Br($W\to e \nu$) = 10.71\%, Br($W\to \mu \nu$) = 10.63\% and 
Br($W\to q q^\prime$)= 67.41\% \cite{Tanabashi:2018oca}.

\subsubsection{$\ttjpsi$ production in tetralepton channel at the LHC}

First, we consider that all the top quarks leptonically decay, which will bring out 
the tetralepton signatures with the process  
$pp\to (t\to W^+(\ell^+\nu)b)
(\bar t \to W^-(\ell^- \bar \nu)\bar b) (\jpsi\to\mu^+\mu^-)$.
All the corresponding possible tetralepton channels
are listed in Tab.\ref{tab:leptonicdecaylist}.
Events should contain two pairs of opposite-sign leptons,
and at least one pair must be mouns.

\begin{table}[h]
\begin{center}
\begin{tabular}{|c|c|c|}
\hline
\hline
 $t\bar{t}$ decay  & $J/\psi$ decay  & Channel  \\
\hline
$(e^+ \nu b)(e^- \bar \nu \bar b)$ & $\mu^+\mu^-$ & $2\mu 2e 2b+ {\slashed E}_T$  \\
$(e^+ \nu b)(\mu^- \bar \nu \bar b)$ & $\mu^+\mu^-$ & $3\mu 1e 2b+ {\slashed E}_T$  \\
$(\mu^+ \nu b)(e^- \bar \nu \bar b)$ & $\mu^+\mu^-$ & $3\mu 1e 2b+ {\slashed E}_T$  \\
$(\mu^+ \nu b)(\mu^- \bar \nu \bar b)$ & $\mu^+\mu^-$ & $4\mu  2b+ {\slashed E}_T$  \\  
\hline
\hline
\end{tabular}
\caption{List of the tetralepton channels for 
$t\bar{t} J/\psi$ production.
}
\label{tab:leptonicdecaylist}
\end{center}
\end{table}

For the channels listed in Tab.\ref{tab:leptonicdecaylist},
all the the final leptons in the signal and background events
satisfy the following cuts:
\bea
 p_{T}^{\ell^\pm} > 2.5\GeV, \,\,\,|\eta_{\ell^\pm}|<2.3, \nn \\
 |m(\mu^+\mu^-)-m_{\jpsi}|<0.5 \GeV, \nn \\
 8.5\GeV< p_{T}^{\jpsi} < 30 \GeV, \,\,\,|y_{\jpsi}|<2.1
 \label{eq:tetra-cut1}
\eea
in the following analyses according to
the Sec.\ref{sec:stable},
where the $\jpsi$ mesons are reconstructed from one pair
of opposite charge mouns.
In case the event includes more than one such moun
pair, the pair with an invariant mass closet to the
nominal value of $m_{\jpsi}$ is attributed to $\jpsi$ mesons
and selected to reconstruct the $\jpsi$ meson candidate.
The two remaning leptons are considered as top quark decay candidates.
The efficiency $\epsilon_b$ we used to correctly tag a $b$-quark jet is approximately 77\%,
as
determined for b-jets with
\be
 p_{T}^{b} > 20\GeV, \,\,\,|\eta_{b}|<2.5
\label{eq:bcut}
\ee
in simulated $t\bar t$ events \cite{btag}.
In this paper, only b-jets statisfy the cuts (\ref{eq:bcut}) are selected.

\begin{table}[h]
\begin{center}
\begin{tabular}{|c|c|ccccc|c|c|}
\hline
\hline
\multirow{2}*{\diagbox{Channels}{Source}}  & \multirow{2}*{Direct} & & & Indirect & & &
\multirow{2}*{Background} & \multirow{2}*{$\cal S$}
\\
\cline{3-7}
& 
& From $ \psi(2S)$
& From $\chi_{c2}$
& From $\chi_{c1}$
& From $\chi_{c0}$ 
& Total Indirect 
&  & \\
\hline
$2\mu 2e 2b+ {\slashed E}_T$ & 0.019  
 & 0.0021	& 0.0035 	& 0.0038  &	0.00005 	& 0.0095 & 0.0015  & 7.4
\\  %

$3\mu 1e 2b+ {\slashed E}_T$ & 0.038  & 0.0042 & 0.0071	& 0.0077	& 0.00010	& 0.019 	& 0.0032  &	10.1
\\  %
$4\mu 2b+ {\slashed E}_T$ & 0.019 & 0.0021	& 0.0035 	& 0.0038 	& 0.00005  	& 0.0095  	& 0.0016	& 7.1 
\\  %
\hline
Combined & 0.076 & 0.0084  &	0.0141  &	0.0153  &	0.00020 &	0.038   & 0.0063 &	14.4
\\  %
\hline
\hline
\end{tabular}
\caption{The cross sections (in unit of fb) for the signal from the direct and prompt $\jpsi$
production 	and background in the tetralepton channels 
at the 13 TeV LHC. The significances  with L = 100 fb$^{-1}$ are also given.
}
\label{tab:tetralep}
\end{center}
\end{table}

In Tab. \ref{tab:tetralep}, we list 
the total cross section for the signal of direct and 
indirect $\jpsi$ production and the SM background in the 
tetralepton channels after considering the cuts 
(\ref{eq:tetra-cut1}) and (\ref{eq:bcut}) and the b-tagging efficiency.
We also list the prompt production $\jpsi$ significance in the last column
with the luminosity L = 100 fb$^{-1}$.
We can see that for the prompt  $\jpsi$ production
in all the tetralepton channels the significances  are all more than 5 with L = 100 fb$^{-1}$.
At the future HL-LHC with L = 3000 fb$^{-1}$, prompt $\jpsi$ production in associated 
with top quark pair can cumulate about more than 300 events in the tetralepton channel.

\subsubsection{$\ttjpsi$ production in trilepton channel at the LHC}

 In this part, we focus on the production rate of $\ttjpsi$
events measured for the final state with three
leptons in the process 
$pp\to (t\to W(q q^\prime) b) ( t \to W(\ell  \nu) b) (\jpsi\to\mu^+\mu^-)$, 
where one top quark  decays hadronically and the
other  leptonically.
We list all the corresponding possible trilepton channels 
in Tab.\ref{tab:semi-leptonicdecaylist}.
Events should contain three leptons,
and at least one pair of opposite-sign mouns.
We also use the cut  (\ref{eq:tetra-cut1}) for the final leptons and
$\jpsi$,  cut  (\ref{eq:bcut}) for final $b$-jets and the b-tagging efficiency
$\epsilon_b=77\%$. In the  $3\mu  2b+ {\slashed E}_T$ channel, 
the opposite-sign moun pair with an invariant mass closet to $m_{\jpsi}$ is
selected to reconstruct the $\jpsi$ meson candidate.
In Tab. \ref{tab:semi-leptonicdecaylist}, we summarize the contributions for the 
total cross section from the signal of  $\jpsi$ produced directly and indirectly
after the selection cuts and the b-tagging efficiency, respectively.
The corresponding irreducible background production rates and the 
prompt $\jpsi$ significances with L = 100 fb$^{-1}$ are also list there.
We can see that the direct contribution for the prompt $\jpsi$ production
in the trilepton channel at 13 TeV LHC can account for about 0.53 fb,
and the indirect contribution is half of the direct.

\begin{table}[h]
\begin{center}
\begin{tabular}{|c|c|c|}
\hline
\hline
 $t\bar{t}$ decay  & $J/\psi$ decay  & Channel  \\
\hline
$(q q^\prime b)(e \nu b)$ & $\mu^+\mu^-$ & $2\mu 1e 2b+ {\slashed E}_T$  \\
$(q q^\prime b)(\mu \nu b)$ & $\mu^+\mu^-$ & $3\mu  2b+ {\slashed E}_T$  \\
\hline
\end{tabular}
\caption{
List of the trilepton channels for 
$t\bar{t} J/\psi$ production.
The
symbols $b$ and $\nu$ denote a b-quark or antiquark and neutrino or antinutrino, respectively,
with charge conjugation implied.
}
\label{tab:semi-leptonicdecaylist}
\end{center}
\end{table}

\begin{table}[h]
\begin{center}
\begin{tabular}{|c|c|ccccc|c|c|}
\hline
\hline
\multirow{2}*{\diagbox{Channels}{Source}}  & \multirow{2}*{Direct} & & & Indirect & & &
\multirow{2}*{Background} & \multirow{2}*{$\cal S$}
\\
\cline{3-7}
& 
& From $ \psi(2S)$
& From $\chi_{c2}$
& From $\chi_{c1}$
& From $\chi_{c0}$ 
& Total Indirect 
&  & \\
\hline
$2\mu 1e 2b+ {\slashed E}_T$ & 0.266 & 0.0292 	& 0.050	& 0.054 &	0.00073  &	0.134 	& 0.020 & 28.3  \\

$3\mu  2b+ {\slashed E}_T$ & 0.264 & 0.0289  & 0.049
& 	0.053  & 	0.00073 & 	0.132 	& 0.021 & 27.3 
\\   %
\hline
Combined & 0.530 & 0.0581 & 0.099	& 0.107  & 	0.00146 & 0.266 & 0.041 & 39.3
\\  %
\hline
\hline
\end{tabular}
\caption{The cross sections (in unit of fb) for the signal from the direct and prompt $\jpsi$
production 	and background in the trilepton channels 
at the 13 TeV LHC. The significances  with L = 100 fb$^{-1}$ are also given.
}
\label{tab:trilep}
\end{center}
\end{table}

\section{Discussion and summary}\label{sec:discussion-and-summary}

In this paper, we investigate the prompt $\jpsi$
production in associated with top quark pair to LO 
in the NRQCD factorization formalism at the 13 TeV LHC.
The prompt $\jpsi$ candidates can be produced directly and indirectly.
The contributions for the indirect $\jpsi$ production
come from radiative decays of $\chi_{cJ}\to \jpsi+\gamma$ 
or hadronic decays of $\psi^\prime\to \jpsi+X$.
We present the total and differential cross section  
for the direct and indirect $\jpsi$ production with the 
basic cut $p_T^{\cal Q} > 3$ GeV. 
We find that the prompt production rates can account for more than 1100 fb and the Fock state
$\sideset{^3}{^{(8)}_1}{\mathop{{S}}}$ gives the dominate contribution to the total and differential cross section.

Then we present the analyses of the prompt $\ttjpsi$ signatures at the LHC.
We consider the $\jpsi$ decaying into a pair of muons, and 
take the $pp\to \ttb \mu^+\mu^-$ process as the irreducible background.
We begin our studies with stable top quarks, and implement the kinematic cuts
for the final reconstructed $\jpsi$ mesons 
with $p_T^{\jpsi} > 8.5$ GeV and $|y_{\jpsi}|<2.1$,
which were used in the ATLAS experiments before.
Through the investigation of the normalized distributions of 
the $p_T^t$, $p_T^{\jpsi}$ and $M_{\ttb}$, we can get 
the maximum significance for the prompt $\jpsi$ production
with addition cuts of  $p_T^{\jpsi} < 30 $ GeV.
In this situation, we can get more than 600 prompt $\ttjpsi$ events
at 13 TeV LHC with L = 100 fb$^{-1}$ via the decay channel $\jpsi\to \mu^+\mu^-$.

Furtherly, we consider the top quarks decay leptonically or hadronically,
and anlysis prompt $\ttb\jpsi$ production in tetralepton and trilepton 
channel at the LHC with the processes 
$pp\to (t\to W^+(\ell^+\nu)b)
(\bar t \to W^-(\ell^- \bar \nu)\bar b) (\jpsi\to\mu^+\mu^-)$
and $pp\to (t\to W(q \bar q)b) ( t \to W(\ell  \nu) b) (\jpsi\to\mu^+\mu^-)$.
In the tetralepton channels, we find that the prompt significance
can all be more than 5 with L = 100 fb$^{-1}$, and at the future
HL-LHC with L = 3000 fb$^{-1}$, the total number of events 
can account for more than 300.
In the trilepton channels, the prompt production  rates
for $\ttb\jpsi$ at 13 TeV LHC can cumulate about 0.53 fb.
We can find that the $\ttjpsi$ production at the LHC have the potential to be detected.
The measurement of the production $\jpsi$ in associated with top quark pair
is useful to investigate the production mechanism of the heavy quarkonium 
and deepen our understanding 
 about the strong interaction.

\acknowledgments
We thank Jinmian Li and Ze-Bo Tang
for helpful discussions. 
This work was supported in part by the National Natural Science 
Foundation of China (Grants No. 11805001,
No. 11305001, No. 11575002, No. 11675033, and No. 11747317)
and the Key Research Foundation of the Education
Ministry of Anhui Province of China (Grant No. KJ2017A032).


\end{document}